\documentclass[10pt,conference,compsocconf]{IEEEtran}
\usepackage{epsfig}
\usepackage{epstopdf}
\usepackage{times}
\usepackage{algorithm}
\usepackage{algorithmic}
\usepackage{amssymb}
\usepackage{amsmath}
\usepackage{subfigure}
\usepackage{multirow}
\usepackage{url}
\usepackage{eso-pic}
\usepackage{lipsum}
\usepackage{soul}
\usepackage{caption}

\usepackage{array} 
\def\ignore#1\endignore{}
\newcolumntype{h}{@{}>{\ignore}l<{\endignore}} 

\newcolumntype{x}[1]{%
>{\centering\hspace{0pt}}p{#1}}%

\usepackage{latexsym}
\usepackage{multicol}
\usepackage{eurosym}
\usepackage{amsthm}
\usepackage{xspace}
\usepackage{pdfpages}
\usepackage{verbatim}

\begin{document}
\title{Context Information for Fast Cell Discovery \\in mm-wave 5G Networks}

\author{
Antonio Capone\IEEEauthorrefmark{1},
Ilario Filippini\IEEEauthorrefmark{1},
Vincenzo Sciancalepore\IEEEauthorrefmark{2}\IEEEauthorrefmark{3}\IEEEauthorrefmark{1}\\
\IEEEauthorrefmark{1} DEIB - Politecnico di Milano \quad\IEEEauthorrefmark{2} IMDEA Networks Institute
\quad\IEEEauthorrefmark{3} Universidad Carlos III de Madrid
\thanks{The research leading to these results has received funding from the European Union's Seventh Framework Program (FP7-ICT-2013-EU-Japan) under grant agreement number 608637 (MiWEBA).}\vspace{-1em}
}

\maketitle
\psfull

\begin{abstract}
The exploitation of the mm-wave bands is one of the most promising solutions for 5G mobile radio networks. However, the use of mm-wave technologies in cellular networks is not straightforward due to mm-wave harsh propagation conditions that limit access availability. In order to overcome this obstacle, hybrid network architectures are being considered where mm-wave small cells can exploit an overlay coverage layer based on legacy technology. The additional mm-wave layer can also take advantage of a functional split between control and user plane, that allows to delegate most of the signaling functions to legacy base stations and to gather context information from users for resource optimization.

However, mm-wave technology requires high gain antenna systems to compensate for high path loss and limited power, e.g., through the use of multiple antennas for high directivity. Directional transmissions must be also used for the cell discovery and synchronization process, and this can lead to a non-negligible delay due to the need to scan the cell area with multiple transmissions at different directions.

In this paper, we propose to exploit the context information related to user position, provided by the separated control plane, to improve the cell discovery procedure and minimize delay. We investigate the fundamental trade-offs of the cell discovery process with directional antennas and the effects of the context information accuracy on its performance. Numerical results are provided to validate our observations. 

\end{abstract}

\section{Introduction}
\label{s:intro}

In this starting phase of the process that will lead to the definition of the 5th generation (5G) of wireless access networks, the mobile industry sector is facing big challenges due to the incredibly fast increase in data traffic volumes and the need to expand the network and provide high peak rates everywhere with low infrastructure costs. Limited resources available in the traditional spectrum portions below 3GHz are pushing the research community to consider much higher frequency bands where large and continuous spectrum segments are available. In particular, millimeter waves (mm-waves) with frequencies above 30GHz are considered excellent candidates for the definition of some of the components of future 5G systems \cite{survey_rappaport,miweba}.

Traditionally, mm-waves have been used for point-to-point radio links and they constitute an important and widely used technology for the backhauling of current mobile networks. For 5G, the main goal is that of enabling the use of mm-waves in the access part of the network so as to enlarge the available capacity and dramatically increase the data rate available for end users. This objective, however, presents a number of technical challenges because of the harsh propagation at so high frequencies and the transmitter technology that limits the power available at antenna elements. Propagation is characterized by severe limitations in the transmission range due to path loss that increases with the square of the frequency. Moreover, most of the objects appear as opaque and cannot be penetrated by signals; also reflection is in most cases very limited depending on the material and the arrival angle. The characterization of the propagation channel is actually a challenge by itself due to the number of different parameters that can influence the signal, and the very diverse physical scenarios to be considered \cite{HHI}.

On the other side, the short wavelength allows the use of antenna arrays with a relatively large number of elements that can be accommodated in a small space, both on the base station (BS) and the mobile terminal (UE). Research in transmission technology is currently focusing on the use of beamforming techniques to concentrate radiated power on very small angles, so as to extend the transmission range, and to track user terminals while they move in the coverage area. This is considered sufficient to enable the design of pico BSs for serving cells with a coverage area up to a few hundreds of meters.

Nevertheless, it is quite evident that even with the massive use of sophisticated directive transmissions the availability of mm-wave access in future mobile networks cannot be continuous either in space, due to limited size of the coverage areas, and in time, due to the obstacles (including human body) that can obstruct the propagation. For this reason, in the MiWEBA project \cite{miweba}, we are considering a system architecture based on heterogeneous technologies where continuous connectivity is provided everywhere using legacy or advanced microwave technologies, while additional capacity layers based on mm-wave small cells are added to provide high transmission rates and traffic offload in the areas with high traffic concentration. The non-continuous coverage of mm-wave small cells can be an issue not only for data traffic, but also for signaling, which in wireless access networks is essential for a number of functions related to mobility management and service access. Therefore, the system architecture that is currently considered by MiWEBA project and other projects on 5G (e.g., MiwaveS \footnote{FP7-ICT EU MiWaveS project http://www.miwaves.eu}) considers a functional split between the user (U) plane and control (C) plane that allows mobile terminal to exchange signaling messages with base stations through legacy technology (typically macro cells) and get radio resources for the data connection on high capacity mm-wave small cells whenever possible \cite{split}.

Obviously, some of the low level control functions that are strictly related to the physical layer cannot be delegated to a separated signaling connection with a different base station. Among these functions,the cell discover in mm-wave access is particularly critical critical, due to the directional transmission techniques adopted during the initial synchronization. Differently from other 3G/4G technologies, where synchronization signals are typically broadcast into the cell without any beamforming scheme, in mm-wave cells it is necessary to adopt more sophisticated techniques. These include some kind of geographical scan of the cell area with directional transmissions. Using highly-directional antennas, UE and BS must sweep through possible angles until they detect each other. This procedure can delay the access to the network, and it can also negatively impact the network behavior during handovers \cite{li2013anchor}.

In this paper, we propose to use the separate C plane to convey to the network some \emph{context} information from UEs that can facilitate the \emph{cell discovery} procedure for the initial synchronization acquisition. In particular, we assume that an external localization service can be used to provide to the network an estimate of the UE position and then use this information to guide the directional transmissions for synchronization. Due to the uncertainty in localization accuracy and in the propagation (that can also occur through reflected rays), the discovery procedure explores a number of beamforming configurations considering the probability to be received by the UE.

The paper is organized as follows. In Section \ref{related} we review previous literature and point out our novel contributions. In Section \ref{s:problem} we present the cell discovery problem, while in Section \ref{s:solutions} we introduce our proposed approach. Section \ref{s:results} includes some numerical results based on simulation, and Section \ref{s:concl} provides concluding remarks.

\section{Related work}
\label{related}

The challenges brought in by the use of directional antennas for C-plane functions have been studied in the past at lower frequency in ad-hoc wireless network scenarios \cite{jakllari2007integrated}.
Considering mm-wave access for Wireless Personal Area Networks (WPANs), devices are assumed to have omnidirectional sensing capabilities, while increasing their directivity towards incoming signals \cite{singh2009blockage}. In addition to that, only 360-degree scanning is used to discover neighbors. In this context,  algorithms for BF tracking \cite{wang2009beam} and route deviation to get around obstacles have been proposed \cite{singh2009blockage}.

More recently, the development of IEEE 802.11ad standard brings mm-wave and directional antennas into Wireless LAN (WLAN) scenarios. According to the standard, transmissions are organized into repeating frames, named Beacon Intervals, and the first part of each frame includes a BF training where an antenna sector sweep is performed. \cite{chen2013directional} proposes a MAC protocol where the antenna pointing direction switches between the best direction found during the sector sweep and a relay station when the channel degrades. The authors in \cite{chandra2014adaptive} present an algorithm for beam-width selection in the contention based period of Beacon Intervals that aims at reaching a subset of users in order to have a high channel occupation and low number of collisions.

The cell discovery problem in mm-wave cellular networks has been considered in \cite{dir_cell_disc_rappaport} and \cite{li2013anchor}. In \cite{li2013anchor}, authors show that there is a mismatch between the area where the network is discoverable (C-plane range) and the area where the mm-wave service is available (U-plane range). In \cite{dir_cell_disc_rappaport}, the cell discovery in directional mm-wave cellular networks is addressed from the physical layer point of view, and the problem of designing the best detector is investigated.

With respect to previous literature, the novel contributions we provide in this paper can be summarized as follow: i) we propose a cell discovery procedure helped by C-/U-plane split, ii) we exploit context information on UE location to guide the search algorithm, iii) we design algorithms that exploit the possibility to select both the beam direction and width, iv) we evaluate the performance of the proposed scheme and compare it with random search and greedy approach. 

\section{mm-Waves Cell Discovery}
\label{s:problem}

The cell discovery procedure in current LTE networks requires the omnidirectional periodic transmission of synchronization signals by the BS. UEs periodically scan the channel to promptly detect the BS and start the synchronization procedure. Due to very high pathloss, omnidirectional transmissions at mm-wave frequencies are not suitable for typical BS-UE distances in cellular networks. Omnidirectional cell discovery in mm-wave systems has the further disadvantage of a big mismatch between the area where the UE can discover a mm-wave BS, very close to the BS itself, and the area where a UE can be served by a mm-wave BS, which extends farther away due to the gain of directional antennas \cite{li2013anchor}. Therefore, the cell discovery process requires directional signaling transmissions throughout different directions to cover the entire service area. Upon beaming a joining UE, tracking and tracing procedures allow the BS to start the association phase with the UE.

A simple strategy for cell discovery could be to sweep through all possible antenna configurations looking for a rendezvous between BS and UE. However, this process might take time and the synchronization process might be delayed. Indeed, with current antenna technologies many different antenna configurations are possible: first, different beam-widths can be available, and second, according to the beam-width, many different pointing directions can be used.

\begin{figure} [!t]
\centering
	\includegraphics[width=0.4\textwidth]{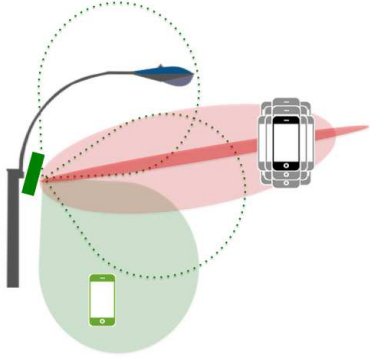}
	\caption{\footnotesize Directional cell discovery examples. }
	\label{fig:dir_search}
	\vspace{-1.5em}
\end{figure}

The beam-width has a strong impact on the discovery procedure. Large beam-widths allow to scan over the surrounding space faster - fewer switches -, but they can reach relatively close UEs. Vice versa, narrow beams can reach far away UEs, but they require a large number of antenna configuration changes to scan the entire space. As a consequence, the discovery procedure must undergo a trade-off between speed and range of the space exploration. Clearly, the best mix depends on available antenna configurations and UE distribution. As shown in Fig.~\ref{fig:dir_search}, the green wider beam allows to explore the space with few pattern switches, however, it can cover only the green UE. The black UE, farther away, cannot be reached, it can be reached with red beam. However, many switches are required to cover the same angular span.

Within the system architecture proposed in MiWEBA, the context information can play a key role in improving the cell discovery delay by providing, at least, information on the positions of BS and UE. However, even richer information can be included, like e.g., channel gain predictions, UE spatial distribution, antenna configurations successfully used in previous accesses, etc. Ideally, if the context provides perfect information, BS and UE could directly point each other with a narrow beam in one step. In practice, some inaccuracy is unavoidable, and the context information can only be used to narrow down the search space. Therefore, some kind of discovery algorithm is required to guide the search through the most likely successful configurations. In the next section we showcase some discovery strategies which we compare considering different levels of context information richness and accuracy. Indeed, the quality of context information may influence the cell discovery algorithm. For instance, Fig.~\ref{fig:dir_search} shows a black UE whose  position information is affected by error; the choice of the algorithm could be either selecting a very narrow and long range beam to increase the gain toward the direction of UE estimated position, or a larger beam-width in order to reach the UE with lower power, but more robust to UE location imprecision.

\begin{figure} [!t]
\centering
	\includegraphics[width=0.4\textwidth]{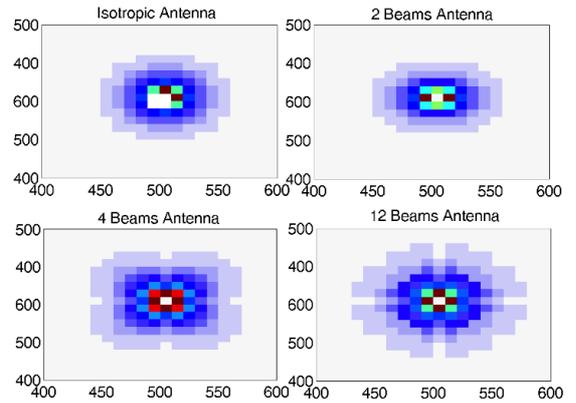}
	\caption{\footnotesize Antenna coverage for different beam-width values depicted on x-y axis, base stations are placed in the middle of the area (point $500$m, $500$m). Received power is averaged over a $10$x$10$m square area. }
	\label{fig:antenna_patterns}
	\vspace{-1.5em}
\end{figure}

In order to boil down the problem to essential aspects, we consider a simplified scenario to which we apply several solutions. We focus on downlink transmissions of the cell synchronization signal, assuming an isotropic antenna for the UE, whereas directional antennas are available at mm-wave BS. We assume a 2D environment (fixed elevation angle) and a steerable antenna, with discrete sets of beam-widths $W_{-3dB}$ and pointing directions $d$. The number of pointing directions is $N=2\pi / W_{-3dB}$. Beam-width and antenna gain are linked by a Gaussian main lobe profile \cite{miwebaD51}:
\begin{equation}
\label{antenna_model}
\renewcommand{\arraystretch}{0.99}
\begin{array}{l}
G_{dB}(\phi,\theta) = \\ 10log\left(\frac{16\pi}{6.76\cdot\phi_{-3dB}\cdot\theta_{-3dB}}\right)-12\cdot\left(\frac{\phi}{\phi_{-3dB}}\right)-12\cdot\left(\frac{\theta}{\theta_{-3dB}}\right) 
\end{array}
\end{equation}
where $\phi$ and $\theta$ are defined as offsets between the main lobe direction and the elevation angle and azimuth angle, correspondingly. Fig.~\ref{fig:antenna_patterns} shows different antenna coverages for different beam-width values and highlights the width-vs-range trade-off.

We cast the cell discovery problem into a time minimization problem where a mm-wave BS must find $i)$ the beam-width $\theta_{-3dB}$, and $ii)$ the pointing direction $d$ in order to provide the UE with a sufficient signal power level to detect the presence of the mm-wave BS, and thus, to synchronize.

\section{Proposed solutions}
\label{s:solutions}

Several solutions to directional neighbor discovery have been already proposed in other fields, such as ad hoc networks and personal area networks, but none of those provides a suitable scheme for future 5G mm-wave network requirements. Interestingly, in this section we show how to cope with the cell discovery problem, focusing our attention on the discovery time, which may become very critical in the 5G network design. 

We thoroughly approach the problem, carrying out a deep analysis on the key parameters presented in the previous section (i.e., beam-width $\theta_{-3dB}$ and pointing direction $d$). This sheds light on how to properly choose those values, in a very short time, in order to provide a sufficient signal level to incoming users to detect the BS presence. 

When a new user joins the network, it starts seeking for a synchronization signal, which is sent through the BS beam.
At one extreme, let us consider no context information is provided when new user joins. BS randomly selects a beam-width and pointing direction and starts probing around to spot potential users. Upon receiving a signal power level above the minimum detection threshold, $Th$, the user reports channel information back and the association process starts. Clearly, this scheme might result in a very long cell discovery time due to the randomness of the user position. In addition, the user distribution heavily influences the discovery time, as the farther users, the thinner the beam-width, the more beam directions must be explored.

At the other extreme, we can assume a perfect knowledge on the context information about user geographical positions. BS readily computes the proper beam-width and the azimuth angle to provide a sufficient signal power level to reach the user. This trivially results in the minimum cell discovery time.

\begin{figure} [!t]
\centering
	\includegraphics[width=0.4\textwidth]{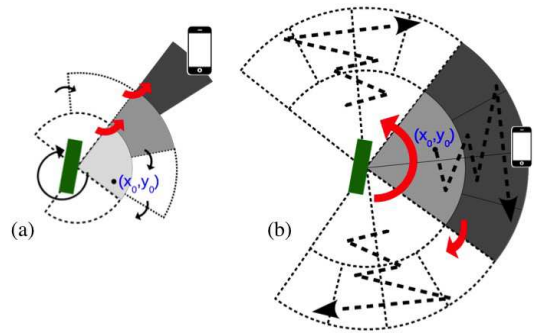}
	\vspace{-2.5em}
	\caption{\footnotesize Illustration of (a) Discovery Greedy Search (b) Enhanced Discovery Procedure. Estimated user location is $(x_0,y_0)$.}
	\label{fig:algo}
	\vspace{-1.5em}
\end{figure}

The real condition is somehow between these two extremes. The cell discovering procedure takes significant advantage of context information regarding the geographical position of the users. As a first improvement to a pure random search, we apply a greedy search paradigm to the process, called \textit{Discovery Greedy Search} (Fig.\ref{fig:algo}(a)). The serving mm-wave BS obtains information regarding the position of incoming users from the macro BS C-plane and properly calculates the pair (beam-width, pointing direction) to reach that position. Note that, a proper beamforming implies the knowledge of the path loss between the user and the BS, which could be estimated by using channel models \cite{HHI} or anchor-based prediction systems \cite{redondi2013context}. However, for the sake of simplicity, we consider here channel path loss and user location errors merged together in a unique equivalent location error. If the user is not detected due to the inaccuracy of the provided position, mm-wave BS starts scanning around through all directions, keeping the same beam-width. If still no user is found, mm-wave BS restarts the circular sweep reducing the beam-width and  iteratively scans the larger set of pointing directions.
When all (beam-width, pointing direction) combinations are explored without any user detection, the user is marked as unreachable.

To smartly exploit the context information, in this paper we propose an enhanced version of the Discovery Greedy Search procedure, named \textit{Enhanced Discovery Procedure} (EDP) (Fig.\ref{fig:algo}(b)). As in the previous approach, when a new user joins the network, the serving BS quickly computes the correct beam-width $\theta^*_{-3dB}$ and pointing direction $d^*$ to properly beam the user, based on the estimated position context information. If the position is not accurate, the user might not be spotted, thus, BS scans the surrounding environment relying on $n$ circular sectors, each $(2\pi/n)$-radian wide. Within the first scanned sector $s$, the sector pointing to the user position, BS starts exploring beam directions adjacent to $d^*$ with a fixed beam-width $\theta^*_{-3dB}$ in order to cover the sector, alternating clockwise and counter-clockwise directions. If no user is reached, BS reduces the beam-width, points the beam again towards the estimated user position, and similarly explores adjacent beam directions within the sector. After completing the scan of the sector without finding the user, the same per-sector beam scanning procedure is repeated for all the other adjacent $n-1$ sectors, alternating clockwise and counter-clockwise sectors. In each sector, the exploration starts from the (beam-width, pointing direction) combination correspondent to ($\theta^*_{-3dB}$,$d^*$) in $s$. The process ends when the UE has been detected or when all $n$ sectors have been probed, resulting in an unreachable user.

The rationale behind the Enhanced Discovery Procedure is to provide a trade-off between two opposite strategies: i) scanning first large azimuthal angles and then extending the range by narrowing the beam, ii) exploring first narrow azimuthal angles until the maximum range is reached and then changing pointing direction. As shown in the next section, the most convenient strategy depends on UE distribution and position accuracy.

\section{Numerical results}
\label{s:results}

We assess the cell discovery time performance by means of numerical simulations. All numerical results are obtained through an ad-hoc MATLAB\textregistered  simulator.
The antenna gain is modelled as a Gaussian main lobe profile described in Eq.~\eqref{antenna_model} while the pathloss model used for the transmission is defined as the following
\begin{equation}
\text{PL} = \alpha + k\cdot 10log\left(\frac{l}{l_0}\right)
\end{equation}
where $\alpha$ is $82.02$dB, reference distance $l_0$ is $5$ meters and $k$ is $2.36$, if the distance between the transmitter and the receiver is longer than the reference distance, or $2.00$ otherwise. For more details, e.g., fading and other channel properties, we refer the reader to~\cite{miwebaD51}. The minimum signal level for PSS acquisition, $Th$, is directly derived from the empirical measurements presented in~\cite{dir_cell_disc_rappaport}, where a Signal-to-Noise-Ratio (SNR) greater than $10$dB has been verified. The smallest selectable beam-width is $0.0157$ rad, which means exploring $360$ non-overlapping beam directions. Larger beam width are obtained by proportionally reducing the number of directions to $180, 120, 90, 72, 60, 48, 24,12, 8, 6, 4, 3, 2$. Test scenarios are provided with $5$ fixed mm-wave BSs with an inter-site distance equal to $200$ meters, and $1000$ users are dropped in the considered area.

\subsection{User distribution impact}
We study the impact of the user distribution on the cell discovery time. We assume no context information about user geographical positions. The BS $i$ chooses a beam-width and a direction (a pair $\{\theta_{-3dB}$,$d\}$) from the set of all possible combinations, according to an uniform distribution. We called this scheme \textit{Random Discovery Algorithm}. Upon a choice, if no user is detected, BS selects another pair of parameters, where each pair can be used only once. If no user is detected after probing all possible pair of values, the user is marked as unreachable. Otherwise, the number of attempts (number of beam-switch operations) performed to detect the user is stored. Note that the number of beam-switches is directly related to the cell discovery time according to the hardware specifications. 

\begin{figure} [!t]
\centering
	\includegraphics[width=0.4\textwidth]{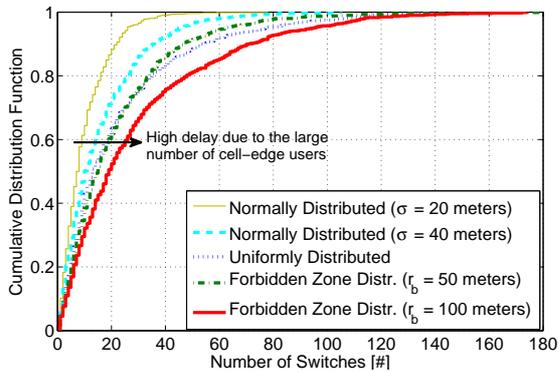}
	\caption{\footnotesize CDF of number of beam switches performed during the cell discovery phase considering different user distributions.}
	\label{fig:cdf_random_search}
	\vspace{-1.5em}
\end{figure}

Fig~\ref{fig:cdf_random_search} shows a cumulative distribution function of the number of beam-switches performed for detecting incoming users in the network. Different user distribution functions are compared in the plot. In the first two, users are dropped in a cell according to a 2D normal distribution with mean value $\mu$ equal to the BS coordinates and standard deviation $\sigma$ equal to $20$ meters and $40$ meters, correspondingly. We consider the same distributions for the last two plots with a slight modification. We create a forbidden zone of $50$ and $100$ meters, respectively, around the center of the cell, where users cannot be placed. This results in placing all the users on the cell-edge which allows us to deeply study the cell-border effect. In the third curve, we drop users in the area according to a uniform random distribution. Results provide useful insights. The cell discovery time is reasonable for a user population uniformly distributed, and even shorter for users close to the mm-wave BSs. If users are far from the BS, the random cell discovery performance degrades, showing a huge number of beam-switches for the $50\%$ of the cases. For those particular scenarios, an improved cell discovery mechanism is needed, as the Random Discovery Algorithm is not reliable.

\subsection{Context-awareness: Geographical Position Accuracy}

We show the performance of context-based algorithms presented in the previous section when the accuracy of position information varies. Fig.~\ref{fig:varySigma_forb} and Fig.~\ref{fig:varyN_forb} refers to users normally distributed with a forbidden zone of 100 meters of radius, while Fig.~\ref{fig:varySigma_norm} and Fig.~\ref{fig:varyN_norm} show results for a normal user distribution.

Fig.~\ref{fig:varySigma_forb} shows the average rendezvous time, as number of antenna configuration switches, when the location accuracy varies. Interesting enough, going from one sector $360$-degree wide to more and narrower sectors allows to reduce the rendezvous time when the location error is reasonably limited. However, when the error increases, the higher error sensibility of narrow sectors severely impacts on the EDP performance, resulting in a more convenient usage of the Greedy procedure. The choice of one $360$-degree sector appears to be the best trade-off for EDP. Results for the Random Discovery Algorithm are not included in the figures, as they do not depend on the location accuracy and show an average rendezvous time of $58.39$ switches, reflecting the infeasibility of such approach.

When the user distribution concentrates users close to the BS, the average number of required switches generally decreases, as shown in results of Fig.~\ref{fig:varySigma_norm} (and Fig.~\ref{fig:varyN_norm}). This is due to the fact that users placed close to the BS can be covered with shorter-range (and wider) beams and, consequently, the whole surrounding area can be scanned in few switches. However, Fig.~\ref{fig:varySigma_norm} reveals an interesting behavior. When the EDP explores $120$-degree and $180$-degree sectors, the rendezvous time increases with respect to results in Fig.~\ref{fig:varySigma_forb}, where users were mainly distributed far from the BS. The motivation behind that is purely geometric: if a close user and a far-away user are characterized by the same location error, the far-away user is detected through a narrow view angle, while the error affecting the close user has a much larger impact on the BS view angle. This implies that location errors for users close to the BS are magnified by angular-selective discovery procedures.

\subsection{Selective Sector Search}

Finally, we assess the performance of the Enhanced Discovery Procedure varying the number of search sectors $n$, e.g., the search sector width.

\begin{figure} [th!]
\centering
	\subfigure[Population of $1000$ users normally distributed with a forbidden zone of 100 meters of radius.]
	{
	\includegraphics[width=0.4\textwidth]{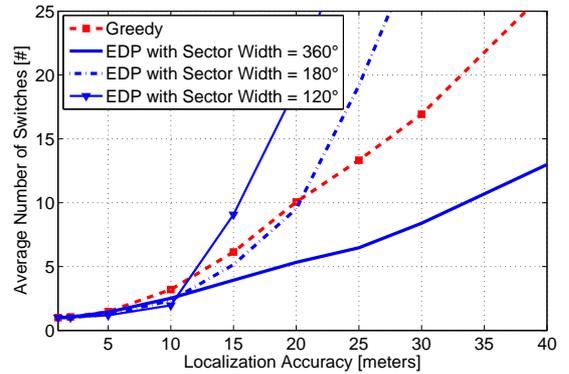}
	\label{fig:varySigma_forb}
	}
	\subfigure[Population of $1000$ users normally distributed with a forbidden zone of 100 meters of radius.]
	{
	\includegraphics[width=0.4\textwidth]{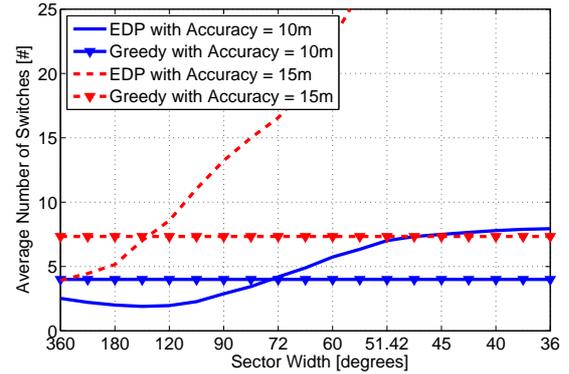}	
	\label{fig:varyN_forb}
	}
	\subfigure[Population of $1000$ users normally distributed.]
	{
	\includegraphics[width=0.4\textwidth]{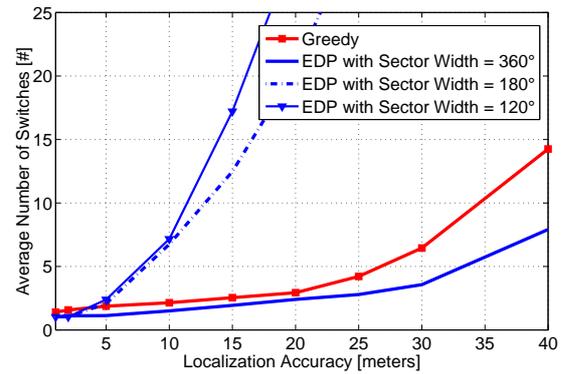}
	\label{fig:varySigma_norm}
	}
	\subfigure[Population of $1000$ users normally distributed.]
	{
	\includegraphics[width=0.4\textwidth]{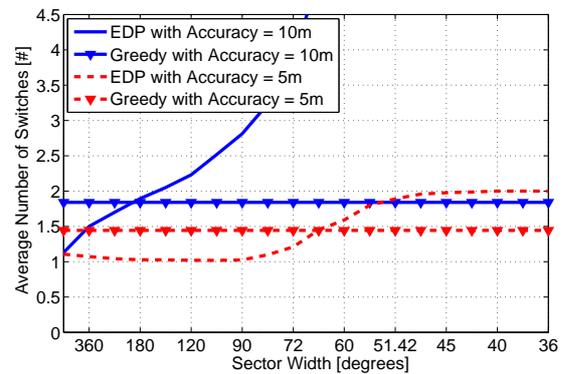}
	\label{fig:varyN_norm}
	}
	\caption{\footnotesize Performance evaluation in terms of average number of beam switches required to detect incoming users for different Search Sector Width and Localization Accuracy values. 
	}
	\label{fig:discovery_algorithms_comp}
\end{figure}

In Fig.~\ref{fig:varyN_forb}, this effect is shown. The figure indicates the average number of antenna configuration switches before a user is discovered using EDP. Results of Discovery Greedy Search are reported as well for the sake of comparison. For $10m$ location accuracy error EDP shows a minimum around $120$ degrees, corresponding to $n=3$. Having fewer and wider sectors slightly increases the rendezvous time, due to the delayed probing of far-away areas, while having more and narrower sectors increases the number of required switches due to the higher sensibility to azimuthal location errors. In such cases, when the number of sectors increases too much, location errors exhibit an adverse effect and the finer EDP exploitation of context is completely cancelled by the sensibility to location errors. In those cases, the Discovery Greedy Search outperforms the Enhanced Discovery Procedure. Additionally, when the location error increases up to and beyond $15m$, the EDP minimum is shifted towards wider sectors, shaping a $360$-degree sector as the best possible choice. 

Fig.~\ref{fig:varyN_norm} analyzes the behavior of the discovery time when varying the number of EDP sectors for normally distributed users. Although it shows a similar behavior to the one verified in Fig.~\ref{fig:varyN_forb}, the impact of the location error is stronger, and thus we show curves for higher accuracy levels.

\section{Conclusions}
\label{s:concl}

We have analyzed the directional cell discovery problem in mm-wave 5G networks characterized by the C-/U-plane split investigated in MiWEBA project. We have shown that currently available solutions are not suitable for mm-wave highly-directional systems and new ones must be designed.

The directional cell discovery issues have been addressed by proposing discovery algorithms enhanced by the context-information available through the separated C-plane link. Results show that the performance of the algorithms depends on user distribution, however, they greatly outperform the random search, even in case of low information accuracy.

We are currently extending the proposed approaches to include more refined context information. Relying on history information, the cell discovery algorithm can learn about user distribution, strong reflection paths or other propagation impairments. They can be exploited to reduce the cell acquisition time by favoring specific antenna configurations.

\section*{Acknowledgments}

The research leading to these results has been partially supported by the EU 7th Framework Program (FP7-ICT-2013-EU-Japan) under grant agreement number 608637 (MiWEBA).

\bibliographystyle{IEEEtran}
\bibliography{bibliography}

\end{document}